\documentclass[reprint,a4paper,amsmath,amssymb,floatfix,showkeys,showpacs]{revtex4-1}

\usepackage{graphicx}
\usepackage{upgreek}

\begin{document}

\title{Ion source for tests of ion behavior in the KATRIN beam line} 

\author{S. Lukic}
\email[(Corresponding author) ]{strahinja.lukic@kit.edu}
\author{B. Bornschein}
\author{G. Drexlin}
\author{F. Gl\"{u}ck}
\altaffiliation{On leave from KFKI, RMKI, Budapest, Hungary}
\author{O. Kazachenko}
\altaffiliation{Present affiliation: ITER Organization, Cadarache, France}
\author{M. C. R. Zoll}
\altaffiliation{Present affiliation: Stockholm University, Sweden}
\affiliation{Karlsruhe Institute of Technology, Karlsruhe, Germany}

\author{M.~Sch\"{o}ppner}
\altaffiliation{Present affiliation: University of Roma Tre, Rome, Italy}
\author{Ch. Weinheimer}
\affiliation{University of M\"{u}nster, M\"{u}nster, Germany}

\date{\today}

\begin{abstract}
An electron-impact ion source based on photoelectron emission was developed for ionization of gases at pressures below $10^{-4} \text{\ mbar}$ in an axial magnetic field in the order of $5 \text{\ T}$. The ion source applies only DC fields, which makes it suitable for use in the presence of equipment sensitive to radio-frequency (RF) fields. The ion source was succesfully tested under varying conditions regarding pressure, magnetic field and magnetic-field gradient, and the results were studied with the help of simulations. The processes in the ion source are well understood and possibilities for further optimization of generated ion currents are clarified.
\end{abstract}


\maketitle 

\begin{figure*}[ht]
\centering
\includegraphics[width=\textwidth]{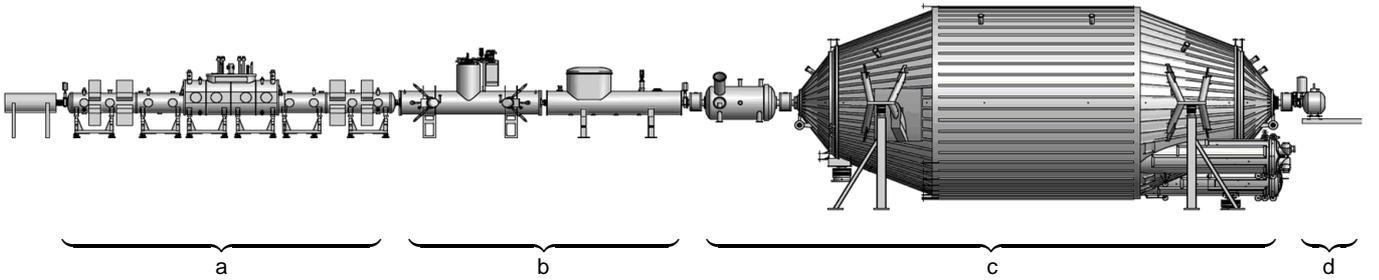}
\caption{\label{Katrin}KATRIN overview: a) tritium source (WGTS), b) differential (DPS2-F) and cryogenic (CPS) pumping sections, c) pre- and main spectrometer, d) detector}
\end{figure*}

\section{Introduction}
\label{}

The KArlsruhe TRItium Neutrino experiment (KATRIN), currently under construction at Campus North of the Karlsruhe Institute of Technology (KIT), will measure the electron neutrino mass in a model-independent way by a kinematic measurement of the shape of the energy spectrum of electrons from tritium $\beta$-decay \cite{KATRIN04}. KATRIN will have a sensitivity on the neutrino mass of $0.2 \text{\ eV/c}^2$, implying an improvement of a factor of ten compared to the precursor experiments at Mainz and Troitsk \cite{Kraus05,Lob99}. This unprecedented sensitivity will enable KATRIN to establish a $5 \sigma$ discovery potential for $m(\nu_e) = 0.35 \text{\ eV/c}^2$. 

Fig. \ref{Katrin} outlines the experimental set-up of KATRIN which adds up to a total length of about 70 m. The $\beta$-electrons in the KATRIN experiment originate from the $10 \text{\ m}$ long Windowless Gaseous Tritium Source (WGTS), followed by differential pumping section (\mbox{DPS2-F}) and a cryogenic pumping section (CPS), which have to reduce the tritium flow rate towards the spectrometer section by 14 orders of magnitude. Superconducting solenoids with magnetic fields up to $5.6 \text{\ T}$ will guide the electrons through vacuum beam-tubes to the high-resolution electrostatic main spectrometer, where the kinetic energy will be measured.

The gaseous tritium source forms a constant density profile with a column density of $5 \times 10^{17} \text{\ cm}^{-2}$. This profile is maintained by continuous injection of $1.8 \text{\ mbar l/s}$ tritium gas into the middle of the WGTS vacuum tube and continuous pumping of the gas at its ends. 

Due to the high activity of the WGTS, ions and ion clusters (e.g. $\text{T}^+, \text{T}_3^{+}, \text{T}_5^{+}, \text{T}^-$) are created in tritium $\beta$-decay itself, as well as in secondary reactions of ions and electrons with tritium molecules. As the endpoint energy of the tritium decay from ionic states is different than from the neutral $\text{T}_2$ molecule, accumulation of ions in the source and the transport section must be prevented in order to avoid distortion of the $\beta$-spectrum. Furthermore, since ions are confined by the magnetic field in a similar way as electrons, they are not effectively pumped by the pumping sections and therefore can reach the spectrometer section. An increased background would be the outcome.

To prevent this, a dipole system for ion removal \cite{Reim09}, as well as two Fourier-Transform Ion Cyclotron Resonance \mbox{(FT-ICR)} ion traps for ion-concentration monitoring \cite{Diaz09} will be installed in the beam line of the \mbox{DPS2-F}. The dipole system uses the $E \times B$ drift to force ions onto a surface with a low-conductivity overlay where they are neutralized. \mbox{FT-ICR} traps rely on analysis of RF signals induced by cyclotron motion of ions to determine their concentration.

In order to test dipoles and \mbox{FT-ICR} traps in conditions that are as close as possible to their KATRIN running conditions, an ion source is needed that can be installed at the entrance of the beam tube to simulate the ion flux that is expected from the WGTS. Moreover, the ion source needs to meet the following requirements:

\begin{itemize}
\item Ionization of gas at pressure in the range $10^{-5} - 10^{-4} \text{\ mbar}$, corresponding to the pressure at the entrance of the \mbox{DPS2-F} beam tube.
\item Operation in an axial magnetic field in the order of $5 \text{\ T}$.
\item Low thermal-energy dissipation suitable for operation in liquid-nitrogen cooled environment.
\item No RF fields, to avoid interference with the \mbox{FT-ICR} system.
\item Production of several $10 \text{\ nA}$ ion current in continuous mode, corresponding to estimated ion flux from the WGTS.
\item Uniform ion current over as large part of the electron-transporting magnetic-flux tube as possible.
\item No production of ions of elements that condense at liquid-nitrogen temperature, in particular metallic ions. Otherwise, such ions could be deposited on top of the low-conductivity overlays of dipoles upon neutralization, and modify their properties. 
\end{itemize}

In the beginning of the study, an extensive literature research was conducted in order to select a suitable ion source for the application described above. A good overview of well-established gas-ionization methods is present in works by Wolf and Brown \cite{Wol95,Bro04}.

Sources employing RF and microwave fields \cite{Con00} were excluded at the outset because of possible interference with \mbox{FT-ICR}. Vacuum discharge ion sources can produce high currents of metallic ions \cite{Wol95}, and are therefore unsuited for the purpose of the test experiment. Photoionization is attractive because light is easy to manipulate, and it is not affected by strong magnetic field. However, because of high photon energies required for ionization of test gases such as deuterium and nitrogen, this mechanism is difficult to apply for this purpose. VUV sources with wavelengths below $80 \text{\ nm}$ are feasible on the laboratory scale \cite{Sch83}, but the required intensities seem to be prohibitive, especially with regards to the low cross sections for photoionization, and the requirement of low gas pressure.

The electron-impact ionization mechanism features cross sections in the order of $10^{-16} \text{\ cm}^{2}$, and high-intensity electron sources are generally feasible. The high magnetic field in \mbox{DPS2-F} is of advantage because it confines electrons radially, thus making their paths longer and increasing the ionization probability. In addition, electrons can be trapped by a suitably formed electrostatic field (Penning trap) for sufficiently long time to make ionization at low pressure effective. For the electron-emitting cathode, the photoelectric emission was selected. It features a negligible heat dissipation, the possibility to illuminate cathodes with relatively large surfaces, as well as full compatibility with strong magnetic fields.  

In the following section the design of the ion source is described, including an initial proof-of-principle test. Experimental setup and results of characterization tests of the ion source are shown in section \ref{Sec-experiment}. Section \ref{Sec-simulation} describes models that were used to simulate particle generation and transport in the ion-source, and compares their results to measured electron and ion currents in tests. The summary and outlook follow in section \ref{Summary}.

\section{Conceptual design of the ion source}
\label{Sec-concept}

\subsection{Electrode-system design}

\begin{figure}
\centering
\includegraphics[width=85mm]{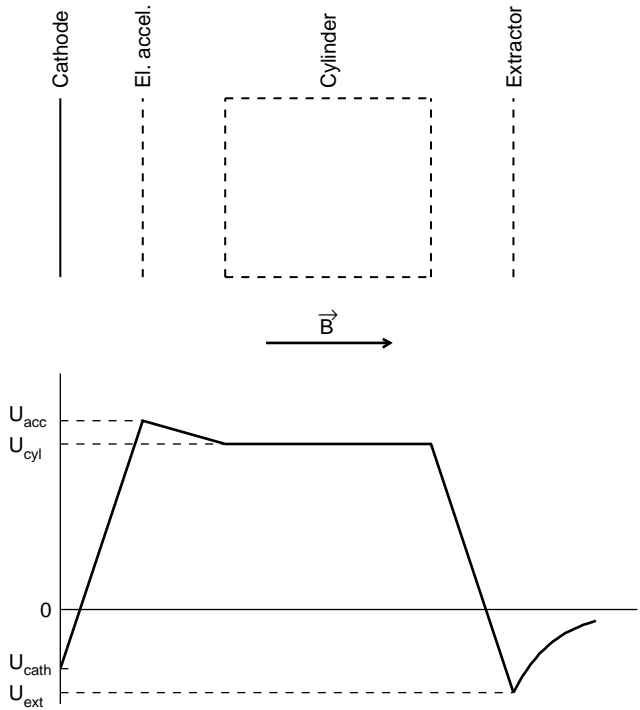}
\caption{\label{electrodes}Electrode configuration of the ion source, with a sketch of the potential distribution along the beam axis}
\end{figure}

In order to maximally profit from the presence of the strong axial magnetic field, the electrode system was designed to not only accelerate electrons to energies suitable for ionization, but to also axially trap the electrons as a variant of Penning trap. Fig. \ref{electrodes} shows a simple electrode geometry sufficient to fulfill this requirement. The magnetic-field direction and a sketch of the potential distribution along the beam axis are also shown. Electrons emitted from the photocathode are accelerated by the potential difference between the electron-acceleration grid and the cathode. The following electrode has a cylindrical shape coaxial with the beam tube, and is kept at slightly lower potential than the acceleration grid in order to screen the ground potential of the surrounding beam tube, thus creating additional volume for ionization. Finally, the ion-extraction grid, at slightly more negative potential than the photocathode, extracts ions in the forward direction, while confining electrons to the space between the photocathode and the ion-extraction grid.\footnote{The actual electrostatic-potential distribution is not trivial in the radial direction, and not constant along the axis in the interior of the cylindrical electrode. However, the simple sketch in fig. \ref{electrodes} represents a close approximation and captures all important features of the real distribution.}

In order to test this concept and to gain insight into ion currents that could be reached, a prototype ion source was built and tested in a simplified setup. 

\subsection{Setup and test of a prototype ion source}

\begin{figure}
\centering
\includegraphics[width=85mm]{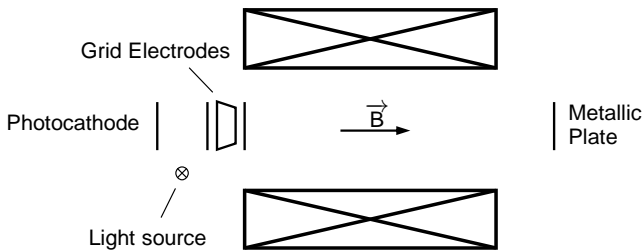}
\caption{\label{SetupPrototype}Sketch of the setup for tests of the prototype ion source}
\end{figure}

The prototype \cite{Schp08} was built with a solid copper cathode irradiated from the front side by a Mercury pen-ray UV lamp LSP035 by Lot-Oriel and tested in a magnetic field of up to 145 mT created by an electromagnet with water cooling. A sketch of the setup is shown in fig.~\ref{SetupPrototype}. Electron and ion currents were measured in a metallic plate downstream from the ion extraction grid. This experimental phase allowed the test of photoemission from copper, as well as stainless-steel cathode materials, confirmed the transport of electrons through the setup and, finally, confirmed the ionization and transport of ions in the downstream direction.

The emission of electrons from the photocathode was tested in high-vacuum conditions ($p < 10^{-6} \text{\ mbar}$) as a function of the electric potential applied to the photocathode. The electron current saturates at $140 \text{\ nA}$ for potential more negative than $-10 \text{\ V}$. Based on irradiance data from Lot-Oriel data sheet for the pen-ray lamp \cite{Lot}, the total number of photons in the $253 \text{\ nm}$ line illuminating the surface of the photocathode was in the order of $10^{16} \text{\ s}^{-1}$, indicating a quantum efficiency in the order of $10^{-4}$.

\begin{figure}
\centering
\includegraphics[width=85mm]{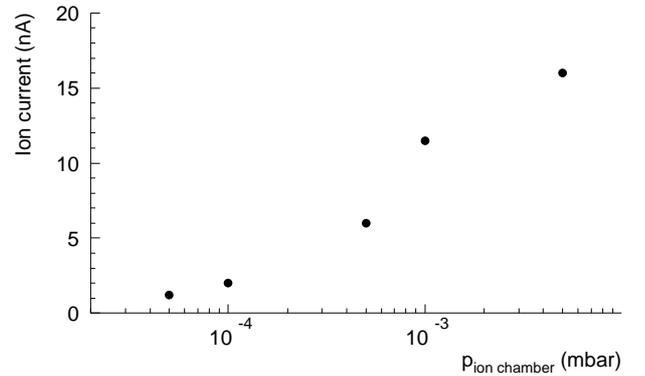}
\caption{\label{ions-p_pt}Ion current measured with the prototype ion source at different pressure levels (data from ref. \citenum{Schp08})}
\end{figure}

In order to measure the ion current, deuterium was injected into the system, and suitable potentials were applied to the ion-source electrodes. The electron-acceleration potential difference was $195 \text{\ V}$ and the magnetic field was $145 \text{\ mT}$. Fig. \ref{ions-p_pt} shows the measured ion currents at different pressure levels. At $5\times10^{-3} \text{\ mbar}$, an ion current of $16 \text{\ nA}$ was measured. At pressures equal or lower than $10^{-4} \text{\ mbar}$, which is more suitable to the test experiment, the measured ion current was between 1 and $2 \text{\ nA}$. This is, however, only part of the total ion production, since in the magnetic field of $145 \text{\ mT}$ the ion movement is not strongly constrained to the magnetic field lines.

These results, approaching the desired ion-current level for the test of the ion-removal and ion-monitoring instrumentation for the \mbox{DPS2-F}, were an encouragement to pursue the development of this ion-source concept one step further. 

\subsection{New design with a back-illuminated cathode}

In a second stage, the geometry of the ion source was adapted for the insertion into the beam tube of \mbox{DPS2-F}. In order to be able to place the light source upstream from the cathode, the cathode was realized as a gold overlay on a fused-silica vacuum window. 

\begin{figure*}
\centering
\includegraphics[width=140mm]{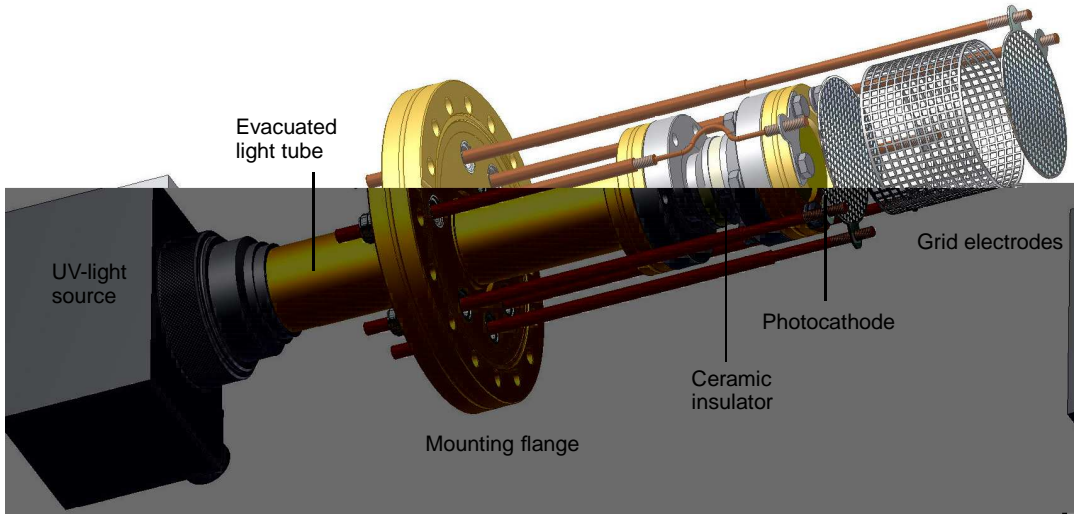}
\caption{\label{Drawing_backlit}Illustration of the ion source with the back-illuminated photocathode. Colors (online only) are intended to help distinguish between details and are not characteristic of the materials used.}
\end{figure*}

An illustration of this version of the ion source is shown in fig. \ref{Drawing_backlit}. A stainless-steel tube (light tube) is welded through the center of a \mbox{CF-100} blind vacuum flange. The light source is mounted onto the rear end of the light tube. On the front end of the light tube, the photocathode is mounted via a ceramic-insulator tube with \mbox{CF-40} vacuum-flange connections. Two contacting bars provide mechanical support and electrical contact to each of the grid electrodes. The contacting bars are made partly of copper and partly of stainless steel, and are mounted onto the central flange via electrical feedthroughs. One additional contacting bar provides the electrical contact to the photocathode via a wire connected to the flange of the photocathode vacuum window. The modular design of the ion source allows adjusting the axial position of electrodes by inserting an additional piece of tubing for the light tube, and using contacting-bar endings of suitable length. When needed, the contacting bars are stabilized against the light tube by mounting a yoke collar made of polyether ether ketone (PEEK).

For the tests described here, a HAMAMATSU L10366 vacuum-sealed deuterium-discharge lamp was used as the light source. Its spectrum features a prominent peak at $160 \text{\ nm}$. To avoid attenuation of light in this frequency range by atmospheric oxygen, the light tube was evacuated down to $\sim 2 \times 10^{-2} \text{\ mbar}$ with a roughing vacuum pump. 

For the photocathode, a fused-silica vacuum window was overlaid with a layer of gold of $20 \text{\ nm}$ thickness, over an intermediate layer of titanium of $10 \text{\ nm}$. The titanium layer provides improved adhesion between the gold layer and the substrate. In order to apply potential to the photocathode, an electrical connection was provided via a feedthrough to the photocathode flange. The electrical contact of the gold overlay with the flange was assured by filling the gap between the fused-silica plate and the flange with vacuum-compatible conductive silver. The useful diameter of the photocathode is $32 \text{\ mm}$, and at distance of $260 \text{\ mm}$ from the light source, it covers about 50\% of the light-spot surface.

\section{Ion-source tests in high magnetic field}
\label{Sec-experiment}

\subsection{Experimental setup}

\begin{figure}
\centering
\includegraphics[width=85mm]{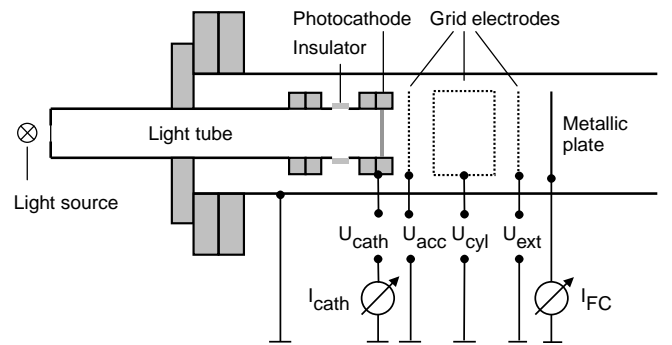}
\caption{\label{Setup_MPIK}Sketch of the test setup at \mbox{MPI-K} Heidelberg with the scheme of electrical connections}
\end{figure}

In order to test the ion source in conditions similar to those expected at \mbox{DPS2-F}, it was installed in the vacuum setup inside the bore of a $4.7 \text{\ T}$ superconducting magnet at the Max-Planck Institute for Nuclear Physics (\mbox{MPI-K}) Heidelberg \cite{Zoll09}. The setup is illustrated in fig. \ref{Setup_MPIK}, including a scheme of electrical connections. A metallic plate was placed immediately behind the ion-extraction electrode to measure electron and ion currents emerging from the ion source. In the ion-current measurements, the potential at the last grid served at the same time as the suppressor electrode to repel back to the plate any secondary electrons created by ion impact. 

\begin{figure}
\centering
\includegraphics[width=85mm]{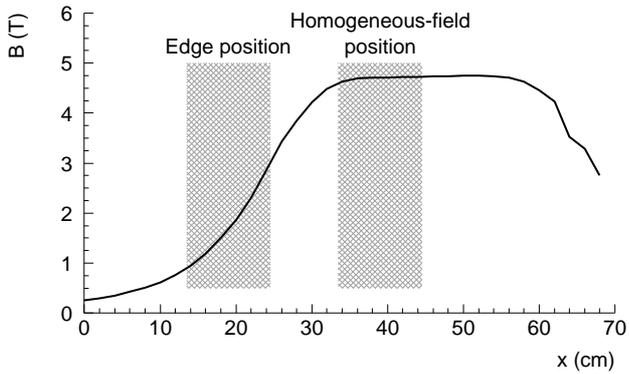}
\caption{\label{B-MPIK}Two test positions of the ion source with respect to the magnetic field}
\end{figure}

All measurements were done for two different axial positions of the ion source, schematically shown against the magnetic-field strength in fig. \ref{B-MPIK}. In the first position, the electrode system was situated at the beginning of the magnet bore, where the magnetic field rises from $0.9 \text{\ T}$ to $3 \text{\ T}$ over the length between the cathode and the extractor electrode. In the second position, the electrodes were situated in the homogeneous magnetic field in the axial center of the coil. Since the ion source is meant to be installed at the entrance of the \mbox{DPS2-F} magnet, tests at the beginning of the magnet bore, where the magnetic-field gradient is somewhat steeper than in the \mbox{DPS2-F}, are of particular interest to reveal possible influence of the magnetic-field gradient on the performance of the source. 

In the following text, the position at the beginning of the magnet bore will be referred to as ``edge position'', and the position in the center of the coil as ``homogeneous-field position''. In the edge position, the light tube is $26 \text{\ cm}$ long and the photocathode is covering 50\% of the light spot of the UV-light source. In the homogeneous-field position, in order to reach the axial center of the coil, the light tube had to be adapted to the length of $46 \text{\ cm}$, so that the photocathode was covering only 16\% of the light spot.

\subsection{Electron current properties}

A standard way to measure photoemission current is to directly measure the current in the photocathode. However, in the present design, as a result of photoemission and ionization processes in the light tube, there are additional components of the total cathode current apart from the electron current emitted into the ionization region. Therefore, the electron emission from the photocathode was tested by measuring the current downstream from the cathode, in the metallic plate shown in fig. \ref{Setup_MPIK}. As in this setup electrons are strongly confined by the magnetic-field lines, all electrons that are not reflected by the magnetic-field gradient or collected on grid electrodes arrive at the metallic plate. The grid structures of the electron-acceleration and the extractor electrodes cover each about 25\% of the cross-sectional area traversed by electrons. The electron-current values shown in the following represent the current measured in the Faraday Cup without correction for electron absorption in the two grids. The pressure in the vacuum tube was kept below $10^{-6} \text{\ mbar}$ during the measurement. The remaining electrodes were at ground potential.

\begin{figure}
\centering
\includegraphics[width=85mm]{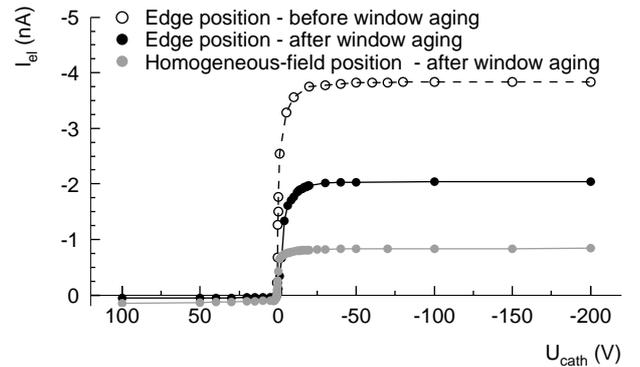}
\caption{\label{e-current}Electron current from the back-illuminated cathode as a function of cathode potential. Two measurement series are shown - before and after cathode window aging under influence of UV light. Data before window aging are taken from ref. \citenum{Zoll09}.}
\end{figure}

Two series of measurements were performed. Between the two series the ion source was in use for roughly 400-500 hours, which caused aging of the UV-lamp window and of the cathode window due to solarization \cite{Key80}. The lamp was replaced before the second series of measurements. In the first series \cite{Zoll09} the ion source was tested only in the edge position, while in the second series both the edge and the homogeneous-field positions were tested. The electron current measured in the metallic plate as a function of cathode potential in both measurement series is shown in fig. \ref{e-current}. In the edge position, the electron current saturates for cathode potentials more negative than $-10 \text{\ V}$. In the first series, the saturation current is $3.8 \text{\ nA}$. In the second series, the saturation current is only $2.0 \text{\ nA}$. This decrease might be caused by aging of the cathode window, but in part also by differences in intensity between individual UV lamps of the same type. In the homogeneous-field position, the electron current was 40\% of that measured in the edge position in the same series. As the geometrical ratio of light-spot coverage in the two positions is only 32\%, this indicates that part of the light that was not directly falling on the photocathode was recovered by reflection from the walls of the light tube. This can be further improved by inserting a high-reflectivity foil into the light tube.

For positive cathode potentials, the electron current in the metallic plate reaches zero at $+2 \text{\ V}$. Due to transmission in the order of 10\% of UV light through the photocathode, the metallic plate is also illuminated and emits electron current of several ten pA, measurable as positive current when the cathode is at positive potential.

\subsection{Ion current properties}

In order to gain good understanding of the performance of the ion source, three tests were performed, in which the ion current was measured 1) as a function of the potential on the cylinder lectrode, 2) as a function of the potential difference between the accelerator electrode and the cathode (electron-acceleration potential) and 3) as a function of pressure. In all measurements, nitrogen was injected as test gas using a flow controller to maintain desired stationary pressure in the system. All results for ion currents presented here refer to measurements after window aging.

In the first test, the effect of the potential applied to the cylinder electrode was studied. In the ion source, ions are created in that part of volume where electrons have energy above ionization threshold, which is true for the largest part of volume between the cathode and the extractor (see fig. \ref{electrodes}). However, ions that are created between the cathode and the electron-acceleration grid are attracted by the electrostatic potential back to the cathode. Only ions that are created downstream from the electron-acceleration grid are accelerated forward and contribute to the ion current in the downstream direction. Applying potential to the cylindrical grid, it is possible to increase the effective ionization volume downstream from the electron-acceleration grid. With increasing cylinder potential up to several volt below the potential on the accelerator grid, the upstream ion current can be expected to drop, while the downstream current rises. 

To test this, the downstream ion current was measured in the metallic plate, and the upstream ion current was determined by measuring the current in the cathode. In order to subtract contributions from the primary electron current and from the currents in the light tube, the cathode current was measured for each point also without gas injection. The difference between the currents measured in the cathode with and without gas injection gives the upstream ion current slightly increased by the contribution of electrons emitted from the cathode by ion impact.

\begin{figure}
\centering
\includegraphics[width=85mm]{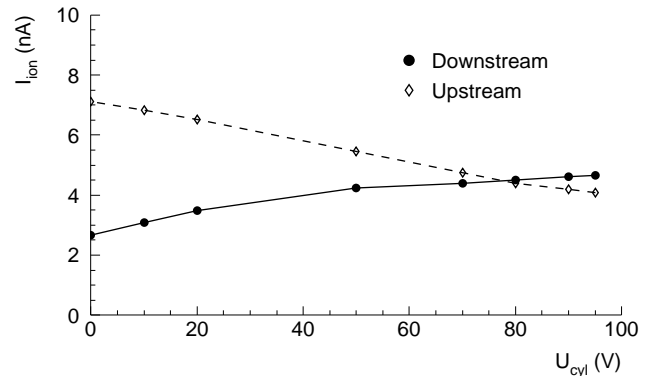}
\caption{\label{ions-fb}Ion currents measured in the upstream and downstream directions as a function of potential on the cylindrical screening electrode, in the edge position ($U_{cath} = -50 \text{\ V}$, $U_{acc}=100 \text{\ V}$, $U_{ext}=-60 \text{\ V}$)}
\end{figure}

With nitrogen injection, the pressure in the system was $1.3 \times 10^{-3} \text{\ mbar}$. The pressure without injection was below $10^{-6} \text{\ mbar}$. The cathode potential was $-50 \text{\ V}$, the electron-accelerator potential was $+100 \text{\ V}$, and the extractor potential was $-60 \text{\ V}$. The upstream and downstream ion currents as a function of the cylinder potential in the edge position are shown in fig. \ref{ions-fb}. As expected, the downstream current increases with increasing cylinder potential, while the upstream current drops. The measured ion currents are higher than the primary-electron current, indicating multiple ionizations per primary electron. This is due to a good containment of electrons and a sufficient electron-acceleration potential to induce secondary processes. In the homogeneous-field position (not shown), the ion current is systematically around 40\% of the corresponding value in the edge position, corresponding well to the ratio of primary electron currents in both positions, despite the difference in the configuration of the magnetic field in the two test positions.

In accordance with the information gained in this measurement, the cylinder potential was kept $10 \text{\ V}$ below the electron-accelerator potential in all subsequent measurements.

\begin{figure}
\centering
\includegraphics[width=85mm]{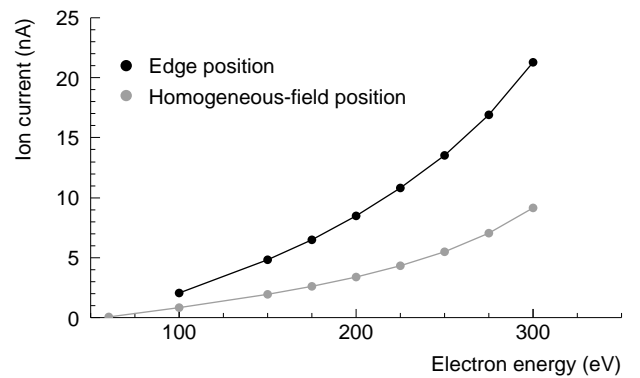}
\caption{\label{ions_vs_acc}Downstream ion current as a function of electron energy, at nitrogen pressure $p = 1.3\times10^{-3} \text{\ mbar}$ ($U_{cath} = -50 \text{\ V}$, $U_{cyl}=U_{acc}-10 \text{\ V}$, $U_{ext}=-60 \text{\ V}$)}
\end{figure}

In the second test, the dependence of the ion current on the electron acceleration was studied. With high acceleration potential, the electrons can be expected to have more energy for multiple ionizations, as well as for creation of secondary electrons capable of ionizing the gas. Fig. \ref{ions_vs_acc} shows the measured ion currents as a function of potential difference between the electron accelerator and the cathode. The ion current rises sharply with electron energy, reaching over five ionizations per primary electron when the electron-acceleration potential is $300 \text{\ V}$ at $p = 1.3\times10^{-3} \text{\ mbar}$. As in the first test, in the homogeneous-field position the ion-current values are systematically 40\% of the corresponding values in the edge position. 

\begin{figure}
\centering
\includegraphics[width=85mm]{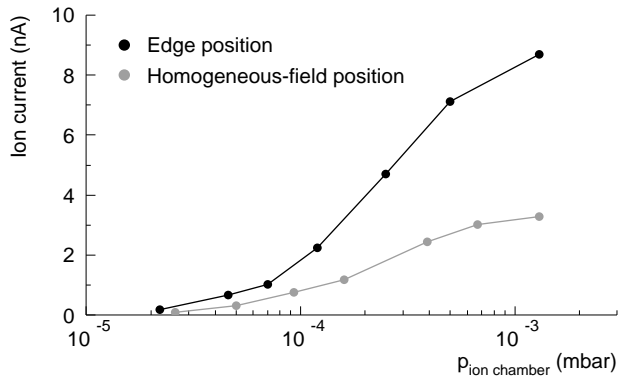}
\caption{\label{ions_vs_p}Ion current as a function of nitrogen pressure in the ion chamber at electron acceleration of 200V ($U_{cath} = -50 \text{\ V}$, $U_{acc}=150 \text{\ V}$, $U_{cyl}=140 \text{\ V}$, $U_{ext}=-60 \text{\ V}$)}
\end{figure}

In the third test, the dependence of the ion current on gas pressure in the ionization volume was studied. Different pressure levels in the ionization volume were achieved by varying the nitrogen-injection rate. In this measurement, the cathode was held at $-50 \text{\ V}$, the electron-accelerator grid at $150 \text{\ V}$, the cylinder at $140 \text{\ V}$, and the extractor at $-60 \text{\ V}$. Fig. \ref{ions_vs_p} shows the measured ion currents as a function of pressure in the ion chamber. As before, the ratio of ion currents in the two test positions is 40\% for all points, corresponding to the ratio of primary-electron currents.

\subsection{Discussion of the experimental results}

Using manufacturer data on angular and spectral distributions of light from the VUV lamp \cite{Ham09}, the total number of photons illuminating the rear surface of the cathode in the spectral range between 150 and $250 \text{\ nm}$ was estimated to be around $4 \times 10^{14}$. After correcting the measured electron current for the absorption in the grid electrodes, one arrives at an overall quantum efficiency of the photocathode of $10^{-4}$. An uncertainty in the order of factor 2 to 3 is present in this estimate because of the uncertainty of the intensity of deuterium lamps, and because the reflection of light in the light tube was neglected. X. Jiang et al.\cite{Jia98} report a quantum efficiency of $10^{-4}$ using a similar photocathode as in this work, illuminated with a $257 \text{\ nm}$ laser, which corresponds to a photon energy just above the threshold for photoelectric emission in gold. In experiments with frontal illumination, Krolikowski and Spicer\cite{Kro70} show that the quantum efficiency increases sharply with increasing photon energy. At a photon energy corresponding to $160 \text{\ nm}$ wavelength, the quantum efficiency in gold is more than one order of magnitude higher than at the threshold, reaching above $10^{-3}$. This energy corresponds to the main peak of the VUV lamp used in this test. The reason why the present photocathode does not reach this efficiency might be in the partial absorption of the $160 \text{\ nm}$ peak by the fused-silica window. This might be improved using a magnesium-fluoride window, characterized by excellent UV transmission down to $120 \text{\ nm}$.

The measured ion current of more than $20 \text{\ nA}$ shows that under suitable conditions multiple ionizations per electron occur, indicating powerful effect of electron trapping on the ionization efficiency. This is crucial for an application like this, where electron-impact ionization is applied at low pressure. The figure of merit is the ion current achievable at pressure of $10^{-4} \text{\ mbar}$. At this moment, this current is $2 \text{\ nA}$. This is roughly the same as the primary-electron current, which is still an excellent result at such low pressure. A very important remark is that in both tested positions the ion current is directly proportional to the primary-electron current at given pressure and at given electrode potentials. This indicates that the magnetic-field gradient, as well as space-charge effects do not significantly alter the properties of electron and ion transport. Thus one can expect that an improvement in primary-electron intensity will result in a proportional increase of the final ion current.

\section{Modeling of the ion source in the final design}
\label{Sec-simulation}

In order to understand the physical processes in the ion source and to optimize its performance, the transport of electrons, as well as their interactions with the gas were simulated using the models described in this section.

\subsection{Transport of electrons}
\label{Sec-e-simulation}

To assess the effect of magnetic mirroring on the transport of electrons, a simple simulation of electron transport was performed in adiabatic approximation. The energy distribution of electrons emitted from the photocathode was sampled from a distribution based on experimental data on photoelectric effect in gold \cite{Kro70}, in conjunction with spectral-intensity data from the manufacturer of the UV-light source \cite{Ham09,Ham06}. The angular distribution of electrons was simulated using basic assumptions of the classical three-step model of photoemission \cite{Ber64}. The inelastic scattering of electrons in metal was not explicitly taken into account. Its effects are partly present in the experimental energy distribution. The space-charge effect was simulated in an iterative way, based on measured saturation electron currents. 

Emission and transport of electrons was simulated under conditions corresponding to the measurement of electron current in the tests of the ion source (see section \ref{Sec-experiment}). 

The validity condition for the adiabatic approximation can be expressed as $\frac{1}{B} \frac{dB}{dt} \ll \omega _{c}$, where $\omega_{c}$ is the cyclotron frequency of the electron, and the expression $\frac{1}{B} \frac{dB}{dt}$ refers to the field seen by the particle in question \cite{Nor63}. This condition is very well fulfilled throughout the setup in both test positions.

The simulated transmission in the homogeneous-field position saturates already for a cathode potential of $-100 \text{\ mV}$, indicating only small space-charge effects. In the edge position, $-6 \text{\ V}$ are sufficient to transmit 99\% of electrons through the magnetic-field gradient. In measured data in both test positions, the electron current rises by another 20\% for higher cathode potentials. This might be due to increased electron reflection from grid electrodes at higher energies, as well as to the production of secondary electrons on the grid, which were both neglected in the simulation. 

\subsection{Ion-current simulation with electron tracking and scattering}

To reproduce ion currents measured in the edge position, field calculations, as well as electron tracking and scattering simulations were performed using the detailed code by one of the authors \cite{Gluck}. The magnetic field was computed using an axisymmetric method with zonal harmonic (Legendre polynomial) expansion; this is much faster than the usual elliptic integral calculation method. As the geometry of the superconducting coil is not available to us, the coil parameters (geometry and amperturns) were adjusted to the measured magnetic field along the axis. For the electric-field computation the boundary element method and an axisymmetric zonal harmonic expansion method were used (the latter is similar to the magnetic field calculation). The grid patterns of the accelerator and extractor grid electrodes are not axisymmetric, but for purpose of field calculation they were approximated by circular ring electrodes. The tracking computation of electrons in the static electric and magnetic fields were made by an 8th order Runge-Kutta code. Both exact trajectory calculation and adiabatic approximation were tested, and found to agree very well. As the adiabatic approximation is much faster, it was used for the final simulations to save computation time. The energy- and angular distributions of electrons at emission were sampled in a similar way as described in section \ref{Sec-e-simulation}.  

Moving from the cathode towards the accelerator grid, the electrons gain kinetic energy with which they traverse the ionization region. Near the extractor grid they are decelerated, and reflected back towards the cathode. Near the accelerator grid the electrons have an energy between 100 and $300 \text{\ eV}$. In this energy region the probability of elastic reflection from metal surfaces is generally small \cite{Cimino04, Harrower56}. Therefore, electrons that hit the grid were considered adsorbed by the grid. Secondary electrons produced in that process have small kinetic energies and are at the minimum of the potential energy in the system, so they were not considered for ionization. Due to the constraint by the magnetic field to move along the field lines, electrons can in principle pass the grid electrodes several times before being adsorbed due to the drift of their magnetron center. Electrons that reach the cathode after traversing the ionization region have an energy below $3 \text{\ eV}$. At such low energy, the interaction of electrons with clean metallic surfaces is characterized by very high elastic-reflection probability \cite{Cimino04, Bauer94, Herlt81, McRae76Ni, McRae76Cu}. Accelerated electrons that, by scattering on gas molecules, lose part of their longitudinal energy that exceeds their initial longitudinal energy, do not reach the cathode any more. Thus electrons often traverse the ionization region several times before being either adsorbed on the cathode, or on the accelerator grid, or until their energy drops below the ionization threshold due to non-elastic collisions. The motion of all secondary electrons created in ionization events was followed as long as they had enough energy for ionization. 

To our knowledge, data for reflection of ultra-low energy electrons on gold surface are unfortunately not available. The simulations presented here were performed with several values for reflection probability. The results are compared and discussed below.

The probability for electron scattering on gas molecules was computed as the product of the path length, the molecule number density and the total scattering cross section. The decision between elastic, electronic excitation, neutral dissociation and ionization collisions was made on the basis of the ratio of the respective cross sections. During a collision, the electron changes its flight-direction angle relative to the magnetic field. At energies relevant to this simulation, the probability for large scattering angles is high. The following experimental cross sections for $\text{electron-N}_2$ scattering were used:
\begin{itemize}
\item{Total and differential elastic cross sections from ref. \citenum{Itikawa}, table 3, and ref. \citenum{Trajmar}, table 25}.
\item{Total electronic excitation and neutral dissociation cross sections from ref. \citenum{Itikawa}, tables 8, 9, 10, 14, and fig. 13; the energy values of the various excitation levels are from table 7 of the same paper; dissociation probabilities of the various electronic excitation levels can be found in refs. \citenum{Zipf} and  \citenum{Cosby}}.
\item{Total ionization cross sections from ref. \citenum{Itikawa}, tables 15-17; for the secondary electron energy distributions tables I.-V. of ref. \citenum{Shyn} were used}.
\end{itemize}
For the angular distributions of inelastic scatterings, cross sections from the code written originally for electron-$\text{H}_2$ scattering \cite{Gluck} were used. Note that the ionization rates in our apparatus are not expected to be sensitive to the electron scattering angle distributions.

Ionization events between the accelerator grid and the extractor grid were counted to obtain the total number $N_{ion}$ of ions that contribute to the downstream ion current. This number was corrected downwards for 25\% absorption probability on the extractor electrode. The ions created by ionization collisions between the cathode and the accelerator grid were not counted as contributing to the downstream ion current, since these ions are accelerated to the cathode, and thus cannot reach the Faraday cup. The secondary electron emission due to the positive ions hitting the cathode was neglected. The ratio of ion current to primary electron current was derived from the ratio of $N_{ion}$ to $N_e$. The primary electron current was obtained from measured saturation electron current (see fig. \ref{e-current}) by correcting for absorption in the two grids between the cathode and the Faraday Cup in the electron-current measurement.  

\begin{figure}
\centering
\includegraphics[width=85mm]{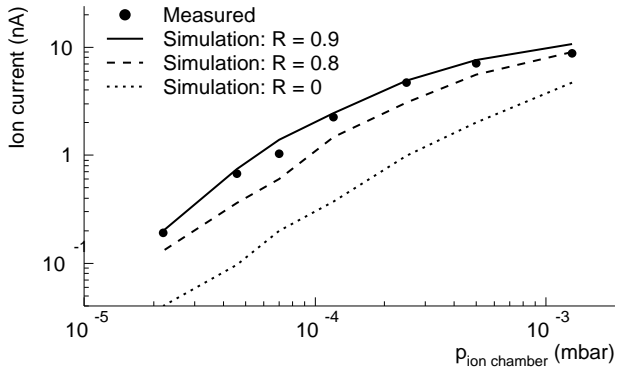}
\caption{\label{i-p-exp-sim}Simulated ion currents at different nitrogen pressures in comparison to measured values. Simulation results for 90\%, 80\%, and zero electron-reflection probability on the cathode surface are shown.}
\end{figure}

Simulations were performed with $N_e=2000$ primary electrons for conditions corresponding to measurements presented in fig. \ref{ions_vs_p} in the edge position. Only for $p = 2.2 \times 10^{-5} \text{\ mbar}$  5000 primary electrons were simulated to reach sufficient statistic. The results are compared with the measured values in fig. \ref{i-p-exp-sim}. Simulation results for 90\%, 80\%, and zero electron-reflection probability on the cathode surface are shown. If the electron reflection from the cathode is neglected, the simulated ion current is roughly proportional to pressure, which is not the case with the measured data. The reflection probability as high as 90\% is required in order to reproduce the shape, as well as the magnitude of the experimental curve. Results at low pressure are particularly sensitive to the used reflection coefficient.

\begin{figure}
\centering
\includegraphics[width=85mm]{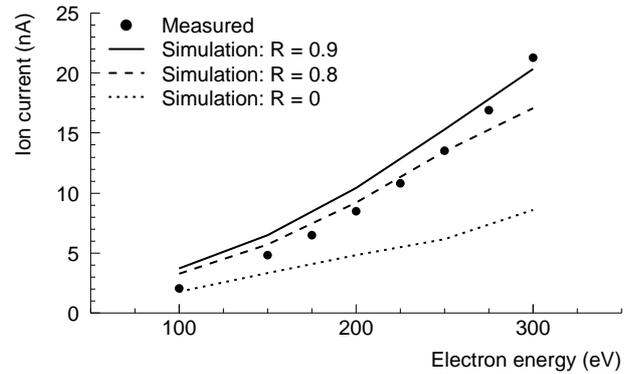}
\caption{\label{i-acc-exp-sim}Simulated ion currents as a function of electron acceleration energy in comparison to measured values. Simulation results for 90\%, 80\%, and zero electron-reflection probability on the cathode surface are shown.}
\end{figure}

Measurements with varying electron energy values (see fig. \ref{ions_vs_acc}) were also simulated. Fig. \ref{i-acc-exp-sim} represents the results for ion currents as a function of electron acceleration energy. Again, simulation results for 90\%, 80\%, and zero electron-reflection probability on the cathode surface are shown. The agreement is somewhat less satisfactory than for the pressure dependence, although the results for 90\% reflection on the cathode are still within a factor of 2 from all measured values. The discrepancy might be due to details of ion transport in the non-uniform magnetic and electrostatic fields, as well as to neglected secondary-electron emission from the cathode when hit by ions created upstream from the accelerator grid.

\section{Summary and outlook}
\label{Summary}

A special type of ion source has been developed to suit stringent requirements of tests on the Transport Section of KATRIN. This ion source is based on electron-impact ionization, whereby electrons are produced by the photoelectric effect. In this way, it is possible to operate the source in moderate as well as high magnetic fields. Due to effective trapping of electrons in the ionization volume, significant ion currents are realized at pressures down to below $10^{-4} \text{\ mbar}$. No RF fields are used in the source, so that it can be applied together with sensitive instrumentation such as \mbox{FT-ICR} that relies on analysis of RF signals induced by cyclotron motion of ions to deduce their concentration. Furthermore, due to low heat dissipation in the high-vacuum part of the volume, the ion source can be used in environment cooled by liquid nitrogen. The concept of the back-illuminated cathode was successfully employed. It has been shown that the ion source can be operated in moderate gradients of magnetic field, such as those that are typically present at the entrance of a superconducting-magnet bore. 

The most important features of the measured ion currents are well reproduced by simulations. Good numerical agreement was achieved after adjusting the, as yet unknown, reflection probability on gold for very-low energy electrons. The value of the reflection probability that allows best reproduction of measured ion currents is plausible in view of experimental results for other metals \cite{Cimino04, Bauer94, Herlt81, McRae76Ni, McRae76Cu}.

It is of high practical interest to increase the ion current at low pressure and moderate ion energies. The most important way to achieve this is by increasing the primary electron current. The quantum efficiency of the present cathode is $10^{-4}$. This can be significantly improved by using a window with better transmission in the deep-UV range. If necessary, a UV-light source with higher intensity can be applied. 

The installation in the \mbox{DPS2-F} beam tube requires that the ion source is configured with the $46 \text{\ cm}$ distance between the light source and the photocathode, implying direct illumination of the cathode by only 16\% of the light spot. The remaining light can be recovered by improving the reflectivity of the light tube. Grid electrodes can be redesigned to reduce absorption of electrons and ions. These are all topics of ongoing optimization work. 

\begin{acknowledgments}

The authors wish to express gratitude to the group for Experimental Techniques of the Institute for Nuclear Physics (IK) at KIT, for highly efficient and competent support with the assembly of both ion sources, as well as with the test setup for the prototype. Particular acknowledgment is due to Mr. Siegfried Horn for key contributions to the mechanical design of the ion source. U. Geckle and Dr. M. Bruns at the Institute for Materials Research III at KIT have kindly provided the titanium-gold overlay for the photocathode of the back-illuminated ion source. Furthermore, we wish to thank the group for stored and cooled ions lead by Prof. Dr. K. Blaum at \mbox{MPI-K} Heidelberg for providing the superconducting magnet for tests in high magnetic field, and for generous technical support on site. Central precision-technical workshop of the \mbox{MPI-K} Heidelberg provided invaluable support on site on more than one occasion.

This work has been funded by the German Ministry for Education and Research under Project codes 05CK5VKA/5, 05A08VK2 and 05CK5PMA/0, as well as by the Deutsche Forschungsgemeinschaft within the frame of the Transregional Collaborative Research Center No. 27 ``Neutrinos and Beyond''.

\end{acknowledgments}

\bibliography{is-arxiv}

%Merlin.mbs v4.21 2009-07-09.
\providecommand{\noopsort}[1]{}\providecommand{\singleletter}[1]{#1}%
\begin{thebibliography}{10}%
\makeatletter
\providecommand \@ifxundefined [1]{%
 \ifx #1\undefined \expandafter \@firstoftwo
 \else \expandafter \@secondoftwo
\fi
}%
\providecommand \@ifnum [1]{%
 \ifnum #1\expandafter \@firstoftwo
 \else \expandafter \@secondoftwo
\fi
}%
\providecommand \enquote [1]{``#1''}%
\providecommand \bibnamefont  [1]{#1}%
\providecommand \bibfnamefont [1]{#1}%
\providecommand \citenamefont [1]{#1}%
\providecommand\href[0]{\@sanitize\@href}%
\providecommand\@href[1]{\endgroup\@@startlink{#1}\endgroup\@@href}%
\providecommand\@@href[1]{#1\@@endlink}%
\providecommand \@sanitize [0]{\begingroup\catcode`\&12\catcode`\#12\relax}%
\@ifxundefined \pdfoutput {\@firstoftwo}{%
 \@ifnum{\z@=\pdfoutput}{\@firstoftwo}{\@secondoftwo}%
}{%
 \providecommand\@@startlink[1]{\leavevmode\special{html:<a href="#1">}}%
 \providecommand\@@endlink[0]{\special{html:</a>}}%
}{%
 \providecommand\@@startlink[1]{%
  \leavevmode
  \pdfstartlink
   attr{/Border[0 0 1 ]/H/I/C[0 1 1]}%
   user{/Subtype/Link/A<</Type/Action/S/URI/URI(#1)>>}%
  \relax
 }%
 \providecommand\@@endlink[0]{\pdfendlink}%
}%
\providecommand \url  [0]{\begingroup\@sanitize \@url }%
\providecommand \@url [1]{\endgroup\@href {#1}{\urlprefix}}%
\providecommand \urlprefix [0]{URL }%
\providecommand \Eprint[0]{\href }%
\@ifxundefined \urlstyle {%
  \providecommand \doi [1]{doi:\discretionary{}{}{}#1}%
}{%
  \providecommand \doi [0]{doi:\discretionary{}{}{}\begingroup
  \urlstyle{rm}\Url }%
}%
\providecommand \doibase [0]{http://dx.doi.org/}%
\providecommand \Doi[1]{\href{\doibase#1}}%
\providecommand \bibAnnote [3]{%
  \BibitemShut{#1}%
  \begin{quotation}\noindent
    \textsc{Key:}\ #2\\\textsc{Annotation:}\ #3%
  \end{quotation}%
}%
\providecommand \bibAnnoteFile [2]{%
  \IfFileExists{#2}{\bibAnnote {#1} {#2} {\input{#2}}}{}%
}%
\providecommand \typeout [0]{\immediate \write \m@ne }%
\providecommand \selectlanguage [0]{\@gobble}%
\providecommand \bibinfo [0]{\@secondoftwo}%
\providecommand \bibfield [0]{\@secondoftwo}%
\providecommand \translation [1]{[#1]}%
\providecommand \BibitemOpen[0]{}%
\providecommand \bibitemStop [0]{}%
\providecommand \bibitemNoStop [0]{.\EOS\space}%
\providecommand \EOS [0]{\spacefactor3000\relax}%
\providecommand \BibitemShut [1]{\csname bibitem#1\endcsname}%
%</preamble>
\bibitem{KATRIN04}%
  \BibitemOpen
  \bibfield{author}{%
  \bibinfo {author} {\bibnamefont{\mbox{KATRIN} Collaboration}},\ }%
  \emph{\bibinfo {title} {\mbox{KATRIN} Design Report 2004}},\ \bibinfo {type}
  {Tech. Rep.}\ (\bibinfo {institution} {Forschungszentrum Karlsruhe,
  Karlsruhe, Germany},\ \bibinfo {year} {2004})%
  \bibAnnoteFile{NoStop}{KATRIN04}%
\bibitem{Kraus05}%
  \BibitemOpen
  \bibfield{author}{%
  \bibinfo {author} {\bibfnamefont{{\relax Ch}.}~\bibnamefont{Kraus}}, \bibinfo
  {author} {\bibfnamefont{B.}~\bibnamefont{Bornschein}}, \bibinfo {author}
  {\bibfnamefont{L.}~\bibnamefont{Bornschein}}, \bibinfo {author}
  {\bibfnamefont{J.}~\bibnamefont{Bonn}}, \bibinfo {author}
  {\bibfnamefont{B.}~\bibnamefont{Flatt}}, \bibinfo {author}
  {\bibfnamefont{A.}~\bibnamefont{Kovalik}}, \bibinfo {author}
  {\bibfnamefont{B.}~\bibnamefont{Ostrick}}, \bibinfo {author}
  {\bibfnamefont{E.~W.}\ \bibnamefont{Otten}}, \bibinfo {author}
  {\bibfnamefont{J.~P.}\ \bibnamefont{Schall}}, \bibinfo {author}
  {\bibfnamefont{{\relax Th}.}~\bibnamefont{Th{\"u}mmler}},\ and\ \bibinfo
  {author} {\bibfnamefont{{\relax Ch}.}~\bibnamefont{Weinheimer}},\ }%
  \bibfield{journal}{%
  \Doi{DOI: 10.1016/S0370-2693(99)00780-7}{\bibinfo {journal} {European
  Physical Journal C}}\ }%
  \textbf{\bibinfo {volume} {40}},\ \bibinfo {pages} {447 } (\bibinfo {year}
  {2005}),\ ISSN \bibinfo {issn} {0370-2693}%
  \bibAnnoteFile{NoStop}{Kraus05}%
\bibitem{Lob99}%
  \BibitemOpen
  \bibfield{author}{%
  \bibinfo {author} {\bibfnamefont{V.~M.}\ \bibnamefont{Lobashev}}, \bibinfo
  {author} {\bibfnamefont{V.~N.}\ \bibnamefont{Aseev}}, \bibinfo {author}
  {\bibfnamefont{A.~I.}\ \bibnamefont{Belesev}}, \bibinfo {author}
  {\bibfnamefont{A.~I.}\ \bibnamefont{Berlev}}, \bibinfo {author}
  {\bibfnamefont{E.~V.}\ \bibnamefont{Geraskin}}, \bibinfo {author}
  {\bibfnamefont{A.~A.}\ \bibnamefont{Golubev}}, \bibinfo {author}
  {\bibfnamefont{O.~V.}\ \bibnamefont{Kazachenko}}, \bibinfo {author}
  {\bibfnamefont{Y.~E.}\ \bibnamefont{Kuznetsov}}, \bibinfo {author}
  {\bibfnamefont{R.~P.}\ \bibnamefont{Ostroumov}}, \bibinfo {author}
  {\bibfnamefont{L.~A.}\ \bibnamefont{Rivkis}}, \bibinfo {author}
  {\bibfnamefont{B.~E.}\ \bibnamefont{Stern}}, \bibinfo {author}
  {\bibfnamefont{N.~A.}\ \bibnamefont{Titov}}, \bibinfo {author}
  {\bibfnamefont{S.~V.}\ \bibnamefont{Zadorozhny}},\ and\ \bibinfo {author}
  {\bibfnamefont{Y.~I.}\ \bibnamefont{Zakharov}},\ }%
  \bibfield{journal}{%
  \Doi{DOI: 10.1016/S0370-2693(99)00781-9}{\bibinfo {journal} {Physics Letters
  B}}\ }%
  \textbf{\bibinfo {volume} {460}},\ \bibinfo {pages} {227 } (\bibinfo {year}
  {1999}),\ ISSN \bibinfo {issn} {0370-2693}%
  \bibAnnoteFile{NoStop}{Lob99}%
\bibitem{Reim09}%
  \BibitemOpen
  \bibfield{author}{%
  \bibinfo {author} {\bibfnamefont{S.}~\bibnamefont{Reimer}},\ }%
  \emph{\bibinfo {title} {Ein elektrostatisches Dipolsystem zur Eliminierung
  von Ionen in der DPS2-F des KATRIN Experimentes}},\ \bibinfo {type} {Diploma
  thesis},\ \bibinfo {school} {Karlsruhe Institute of Technology} (\bibinfo
  {year} {2009})%
  \bibAnnoteFile{NoStop}{Reim09}%
\bibitem{Diaz09}%
  \BibitemOpen
  \bibfield{author}{%
  \bibinfo {author} {\bibfnamefont{M.}~\bibnamefont{Ubieto-D\'iaz}}, \bibinfo
  {author} {\bibfnamefont{D.}~\bibnamefont{Rodr\'iguez}}, \bibinfo {author}
  {\bibfnamefont{S.}~\bibnamefont{Lukic}}, \bibinfo {author}
  {\bibfnamefont{{\relax Sz}.}~\bibnamefont{Nagy}}, \bibinfo {author}
  {\bibfnamefont{S.}~\bibnamefont{Stahl}},\ and\ \bibinfo {author}
  {\bibfnamefont{K.}~\bibnamefont{Blaum}},\ }%
  \bibfield{journal}{%
  \Doi{DOI: 10.1016/j.ijms.2009.07.003}{\bibinfo {journal} {International
  Journal of Mass Spectrometry}}\ }%
  \textbf{\bibinfo {volume} {288}},\ \bibinfo {pages} {1 } (\bibinfo {year}
  {2009}),\ ISSN \bibinfo {issn} {1387-3806},\
  \Eprint{http://arxiv.org/abs/0907.3458}{arXiv:0907.3458}%
  \bibAnnoteFile{NoStop}{Diaz09}%
\bibitem{Wol95}%
  \BibitemOpen
  \bibfield{author}{%
  \bibinfo {author} {\bibfnamefont{B.}~\bibnamefont{Wolf}},\ }%
  \emph{\bibinfo {title} {Handbook of Ion Sources}}\ (\bibinfo {publisher}
  {CRC, Boca Raton},\ \bibinfo {year} {1995})%
  \bibAnnoteFile{NoStop}{Wol95}%
\bibitem{Bro04}%
  \BibitemOpen
  \bibfield{author}{%
  \bibinfo {author} {\bibfnamefont{I.~G.}\ \bibnamefont{Brown}},\ }%
  \emph{\bibinfo {title} {The Physics and Technology of Ion Sources}}\
  (\bibinfo {publisher} {Wiley-VCH},\ \bibinfo {year} {2004})%
  \bibAnnoteFile{NoStop}{Bro04}%
\bibitem{Con00}%
  \BibitemOpen
  \bibfield{author}{%
  \bibinfo {author} {\bibfnamefont{H.}~\bibnamefont{Conrads}}\ and\ \bibinfo
  {author} {\bibfnamefont{M.}~\bibnamefont{Schmidt}},\ }%
  \bibfield{journal}{%
  \bibinfo {journal} {Plasma Sources Science and Technology}\ }%
  \textbf{\bibinfo {volume} {9}},\ \bibinfo {pages} {441} (\bibinfo {year}
  {2000})%
  \bibAnnoteFile{NoStop}{Con00}%
\bibitem{Sch83}%
  \BibitemOpen
  \bibfield{author}{%
  \bibinfo {author} {\bibfnamefont{G.}~\bibnamefont{Sch\"onhense}}\ and\
  \bibinfo {author} {\bibfnamefont{U.}~\bibnamefont{Heinzmann}},\ }%
  \bibfield{journal}{%
  \bibinfo {journal} {Journal of Physics E: Scientific Instruments}\ }%
  \textbf{\bibinfo {volume} {16}},\ \bibinfo {pages} {74} (\bibinfo {year}
  {1983})%
  \bibAnnoteFile{NoStop}{Sch83}%
\bibitem{Note1}%
  \BibitemOpen
  \bibinfo {note} {The actual electrostatic-potential distribution is not
  trivial in the radial direction, and not constant along the axis in the
  interior of the cylindrical electrode. However, the simple sketch in fig.
  \ref {electrodes} represents a close approximation and captures all important
  features of the real distribution.}%
  \bibAnnoteFile{Stop}{Note1}%
\bibitem{Schp08}%
  \BibitemOpen
  \bibfield{author}{%
  \bibinfo {author} {\bibfnamefont{M.}~\bibnamefont{Sch{\"o}ppner}},\ }%
  \emph{\bibinfo {title} {A Prototype Ion Source for the Functionality Test of
  the KATRIN Transport Section}},\ \bibinfo {type} {Diploma thesis},\ \bibinfo
  {school} {Westf{\"a}lische Wilhelms-Universit\"at M{\"u}nster} (\bibinfo
  {year} {2008})%
  \bibAnnoteFile{NoStop}{Schp08}%
\bibitem{Lot}%
  \BibitemOpen
  \bibinfo {organization} {Lot-Oriel Group Europe},\ \emph{\bibinfo {title}
  {Lamp spectra and irradiance data}},\ \bibinfo {note} {www.lot-oriel.com}%
  \bibAnnoteFile{NoStop}{Lot}%
\bibitem{Zoll09}%
  \BibitemOpen
  \bibfield{author}{%
  \bibinfo {author} {\bibfnamefont{M.~C.~R.}\ \bibnamefont{Zoll}},\ }%
  \emph{\bibinfo {title} {Development of tools and methods for KATRIN DPS2-F
  test experiments}},\ \bibinfo {type} {Diploma thesis},\ \bibinfo {school}
  {University of Karlsruhe} (\bibinfo {year} {2009})%
  \bibAnnoteFile{NoStop}{Zoll09}%
\bibitem{Key80}%
  \BibitemOpen
  \bibfield{author}{%
  \bibinfo {author} {\bibfnamefont{P.~J.}\ \bibnamefont{Key}}\ and\ \bibinfo
  {author} {\bibfnamefont{R.~C.}\ \bibnamefont{Preston}},\ }%
  \bibfield{journal}{%
  \bibinfo {journal} {Journal of Physics E: Scientific Instruments}\ }%
  \textbf{\bibinfo {volume} {13}},\ \bibinfo {pages} {866} (\bibinfo {year}
  {1980})%
  \bibAnnoteFile{NoStop}{Key80}%
\bibitem{Ham09}%
  \BibitemOpen
  \bibinfo {note} {Data provided to the authors for information purposes by
  HAMAMATSU Photonics Deutschland GmbH}%
  \bibAnnoteFile{NoStop}{Ham09}%
\bibitem{Jia98}%
  \BibitemOpen
  \bibfield{author}{%
  \bibinfo {author} {\bibfnamefont{X.}~\bibnamefont{Jiang}}, \bibinfo {author}
  {\bibfnamefont{C.~N.}\ \bibnamefont{Berglund}}, \bibinfo {author}
  {\bibfnamefont{A.~E.}\ \bibnamefont{Bell}},\ and\ \bibinfo {author}
  {\bibfnamefont{W.~A.}\ \bibnamefont{Mackie}},\ }%
  \bibfield{journal}{%
  \Doi{10.1116/1.590462}{\bibinfo {journal} {J. Vac. Sci. Technol.}}\ }%
  \textbf{\bibinfo {volume} {B16}},\ \bibinfo {pages} {3374} (\bibinfo {year}
  {1998})%
  \bibAnnoteFile{NoStop}{Jia98}%
\bibitem{Kro70}%
  \BibitemOpen
  \bibfield{author}{%
  \bibinfo {author} {\bibfnamefont{W.~F.}\ \bibnamefont{Krolikowski}}\ and\
  \bibinfo {author} {\bibfnamefont{W.~E.}\ \bibnamefont{Spicer}},\ }%
  \bibfield{journal}{%
  \Doi{10.1103/PhysRevB.1.478}{\bibinfo {journal} {Phys. Rev. B}}\ }%
  \textbf{\bibinfo {volume} {1}},\ \bibinfo {pages} {478} (\bibinfo {month}
  {Jan}\ \bibinfo {year} {1970})%
  \bibAnnoteFile{NoStop}{Kro70}%
\bibitem{Ham06}%
  \BibitemOpen
  \bibinfo {organization} {HAMAMATSU Photonics},\ \emph{\bibinfo {title}
  {High-brightness VUV light-source unit - L10366 series}},\ \bibinfo {note}
  {www.hamamatsu.com}%
  \bibAnnoteFile{NoStop}{Ham06}%
\bibitem{Ber64}%
  \BibitemOpen
  \bibfield{author}{%
  \bibinfo {author} {\bibfnamefont{C.~N.}\ \bibnamefont{Berglund}}\ and\
  \bibinfo {author} {\bibfnamefont{W.~E.}\ \bibnamefont{Spicer}},\ }%
  \bibfield{journal}{%
  \Doi{10.1103/PhysRev.136.A1030}{\bibinfo {journal} {Phys. Rev.}}\ }%
  \textbf{\bibinfo {volume} {136}},\ \bibinfo {pages} {A1030} (\bibinfo {month}
  {Nov}\ \bibinfo {year} {1964})%
  \bibAnnoteFile{NoStop}{Ber64}%
\bibitem{Nor63}%
  \BibitemOpen
  \bibfield{author}{%
  \bibinfo {author} {\bibfnamefont{T.~G.}\ \bibnamefont{Northrop}},\ }%
  \emph{\bibinfo {title} {On adiabatic motion of charged particles}}\ (\bibinfo
  {publisher} {New York: Interscience},\ \bibinfo {year} {1963})%
  \bibAnnoteFile{NoStop}{Nor63}%
\bibitem{Gluck}%
  \BibitemOpen
  \bibfield{author}{%
  \bibinfo {author} {\bibfnamefont{F.}~\bibnamefont{Gl\"uck}},\ }%
  \bibinfo {howpublished} {to be published}%
  \bibAnnoteFile{NoStop}{Gluck}%
\bibitem{Cimino04}%
  \BibitemOpen
  \bibfield{author}{%
  \bibinfo {author} {\bibfnamefont{R.}~\bibnamefont{Cimino}}, \bibinfo {author}
  {\bibfnamefont{I.~R.}\ \bibnamefont{Collins}}, \bibinfo {author}
  {\bibfnamefont{M.~A.}\ \bibnamefont{Furman}}, \bibinfo {author}
  {\bibfnamefont{M.}~\bibnamefont{Pivi}}, \bibinfo {author}
  {\bibfnamefont{F.}~\bibnamefont{Ruggiero}}, \bibinfo {author}
  {\bibfnamefont{G.}~\bibnamefont{Rumolo}},\ and\ \bibinfo {author}
  {\bibfnamefont{F.}~\bibnamefont{Zimmermann}},\ }%
  \bibfield{journal}{%
  \Doi{10.1103/PhysRevLett.93.014801}{\bibinfo {journal} {Phys. Rev. Lett.}}\
  }%
  \textbf{\bibinfo {volume} {93}},\ \bibinfo {pages} {014801} (\bibinfo {year}
  {2004})%
  \bibAnnoteFile{NoStop}{Cimino04}%
\bibitem{Harrower56}%
  \BibitemOpen
  \bibfield{author}{%
  \bibinfo {author} {\bibfnamefont{G.~A.}\ \bibnamefont{Harrower}},\ }%
  \bibfield{journal}{%
  \Doi{10.1103/PhysRev.104.52}{\bibinfo {journal} {Phys. Rev.}}\ }%
  \textbf{\bibinfo {volume} {104}},\ \bibinfo {pages} {52} (\bibinfo {year}
  {1956})%
  \bibAnnoteFile{NoStop}{Harrower56}%
\bibitem{Bauer94}%
  \BibitemOpen
  \bibfield{author}{%
  \bibinfo {author} {\bibfnamefont{E.}~\bibnamefont{Bauer}},\ }%
  \bibfield{journal}{%
  \bibinfo {journal} {Reports on Progress in Physics}\ }%
  \textbf{\bibinfo {volume} {57}},\ \bibinfo {pages} {895} (\bibinfo {year}
  {1994})%
  \bibAnnoteFile{NoStop}{Bauer94}%
\bibitem{Herlt81}%
  \BibitemOpen
  \bibfield{author}{%
  \bibinfo {author} {\bibfnamefont{H.~J.}\ \bibnamefont{Herlt}}, \bibinfo
  {author} {\bibfnamefont{R.}~\bibnamefont{Feder}}, \bibinfo {author}
  {\bibfnamefont{G.}~\bibnamefont{Meister}},\ and\ \bibinfo {author}
  {\bibfnamefont{E.~G.}\ \bibnamefont{Bauer}},\ }%
  \bibfield{journal}{%
  \Doi{DOI: 10.1016/0038-1098(81)90790-0}{\bibinfo {journal} {Solid State
  Communications}}\ }%
  \textbf{\bibinfo {volume} {38}},\ \bibinfo {pages} {973 } (\bibinfo {year}
  {1981}),\ ISSN \bibinfo {issn} {0038-1098}%
  \bibAnnoteFile{NoStop}{Herlt81}%
\bibitem{McRae76Ni}%
  \BibitemOpen
  \bibfield{author}{%
  \bibinfo {author} {\bibfnamefont{E.}~\bibnamefont{McRae}}\ and\ \bibinfo
  {author} {\bibfnamefont{C.}~\bibnamefont{Caldwell}},\ }%
  \bibfield{journal}{%
  \Doi{DOI: 10.1016/0039-6028(76)90168-0}{\bibinfo {journal} {Surface
  Science}}\ }%
  \textbf{\bibinfo {volume} {57}},\ \bibinfo {pages} {63 } (\bibinfo {year}
  {1976}),\ ISSN \bibinfo {issn} {0039-6028}%
  \bibAnnoteFile{NoStop}{McRae76Ni}%
\bibitem{McRae76Cu}%
  \BibitemOpen
  \bibfield{author}{%
  \bibinfo {author} {\bibfnamefont{E.}~\bibnamefont{McRae}}\ and\ \bibinfo
  {author} {\bibfnamefont{C.}~\bibnamefont{Caldwell}},\ }%
  \bibfield{journal}{%
  \Doi{DOI: 10.1016/0039-6028(76)90169-2}{\bibinfo {journal} {Surface
  Science}}\ }%
  \textbf{\bibinfo {volume} {57}},\ \bibinfo {pages} {77 } (\bibinfo {year}
  {1976}),\ ISSN \bibinfo {issn} {0039-6028}%
  \bibAnnoteFile{NoStop}{McRae76Cu}%
\bibitem{Itikawa}%
  \BibitemOpen
  \bibfield{author}{%
  \bibinfo {author} {\bibfnamefont{Y.}~\bibnamefont{Itikawa}},\ }%
  \bibfield{journal}{%
  \bibinfo {journal} {J. Phys. Chem. Ref. Data}\ }%
  \textbf{\bibinfo {volume} {35}},\ \bibinfo {pages} {31} (\bibinfo {year}
  {2006})%
  \bibAnnoteFile{NoStop}{Itikawa}%
\bibitem{Trajmar}%
  \BibitemOpen
  \bibfield{author}{%
  \bibinfo {author} {\bibfnamefont{S.}~\bibnamefont{Trajmar}}, \bibinfo
  {author} {\bibfnamefont{D.~F.}\ \bibnamefont{Register}},\ and\ \bibinfo
  {author} {\bibfnamefont{A.}~\bibnamefont{Chutjian}},\ }%
  \bibfield{journal}{%
  \Doi{DOI: 10.1016/0370-1573(83)90071-6}{\bibinfo {journal} {Physics
  Reports}}\ }%
  \textbf{\bibinfo {volume} {97}},\ \bibinfo {pages} {219 } (\bibinfo {year}
  {1983})%
  \bibAnnoteFile{NoStop}{Trajmar}%
\bibitem{Zipf}%
  \BibitemOpen
  \bibfield{author}{%
  \bibinfo {author} {\bibfnamefont{E.}~\bibnamefont{Zipf}}\ and\ \bibinfo
  {author} {\bibfnamefont{R.}~\bibnamefont{McLaughlin}},\ }%
  \bibfield{journal}{%
  \Doi{DOI: 10.1016/0032-0633(78)90066-1}{\bibinfo {journal} {Planetary and
  Space Science}}\ }%
  \textbf{\bibinfo {volume} {26}},\ \bibinfo {pages} {449 } (\bibinfo {year}
  {1978})%
  \bibAnnoteFile{NoStop}{Zipf}%
\bibitem{Cosby}%
  \BibitemOpen
  \bibfield{author}{%
  \bibinfo {author} {\bibfnamefont{P.~C.}\ \bibnamefont{Cosby}},\ }%
  \bibfield{journal}{%
  \Doi{10.1063/1.464385}{\bibinfo {journal} {The Journal of Chemical Physics}}\
  }%
  \textbf{\bibinfo {volume} {98}},\ \bibinfo {pages} {9544} (\bibinfo {year}
  {1993})%
  \bibAnnoteFile{NoStop}{Cosby}%
\bibitem{Shyn}%
  \BibitemOpen
  \bibfield{author}{%
  \bibinfo {author} {\bibfnamefont{T.~W.}\ \bibnamefont{Shyn}},\ }%
  \bibfield{journal}{%
  \Doi{10.1103/PhysRevA.27.2388}{\bibinfo {journal} {Phys. Rev. A}}\ }%
  \textbf{\bibinfo {volume} {27}},\ \bibinfo {pages} {2388} (\bibinfo {year}
  {1983})%
  \bibAnnoteFile{NoStop}{Shyn}%
\end{thebibliography}%

\end{document}